\documentclass[onecolumn,authoryear]{els-mrw} 

\usepackage{amsmath,amssymb,amsfonts,amsthm,makeidx,graphicx}
\usepackage{txfonts}
\usepackage{helvet}

\begin{document}

\chapter[Semi-empirical models]{Semi-empirical Models of Galaxy Formation and Evolution}\label{chap1}

\author[1,2,3,4]{Andrea Lapi}
\author[1,2,5]{Lumen Boco}
\author[6]{Francesco Shankar}
\address[1]{\orgname{Scuola Internazionale Superiore Studi Avanzati (SISSA)}, \orgdiv{Physics Area}, \orgaddress{Via Bonomea 265, 34136 Trieste, Italy}}
\address[2]{\orgname{Institute for Fundamental Physics of the Universe (IFPU)}, \orgaddress{Via Beirut 2, 34014 Trieste}}
\address[3]{\orgname{Istituto Nazionale Fisica Nucleare (INFN)}, \orgdiv{Sezione di Trieste}, \orgaddress{ Via Valerio 2, 34127 Trieste, Italy}}
\address[4]{\orgname{Istituto di Radio-Astronomia (IRA-INAF)}, \orgaddress{Via Gobetti 101, 40129 Bologna, Italy}}
\address[5]{\orgname{Institut fur Theoretische Astrophysik, ZAH, Universitat Heidelberg, Albert-Ueberle-Str. 3, 69120 Heidelberg, Germany}}
\address[6]{\orgname{School of Physics and Astronomy, University of Southampton, Highfield, Southampton, SO17 1BJ, UK}}

\articletag{Chapter Article}

\maketitle

\begin{glossary}[Glossary]
\term{Galaxies} gravitationally bound ensembles of stars, gas and dust found throughout the universe.\\
\term{Dark matter} putative non-luminous matter interacting with the ordinary one (e.g., atoms) solely via gravitational forces. \\
\term{Black holes} celestial objects featuring a gravitational field so strong that even light cannot escape it.\\
\end{glossary}

\begin{glossary}[Nomenclature]
\begin{tabular}{@{}lp{34pc}@{}}
DM & Dark Matter\\
SIMs & (Hydrodynamical) Simulations\\
SAMs & Semi Analytic Models\\
SEMs & Semi Empirical Models\\
SMHM & Stellar Mass Halo Mass (Relationship)\\
(s)SFR & (Specific) Star Formation Rate\\
(s)HAR & (Specific) Halo Accretion Rate\\
BHs & Black Holes\\
\end{tabular}
\end{glossary}

\begin{abstract}[Abstract]
We provide a review on semi-empirical models of galaxy formation and evolution. In Section \ref{sec|intro} we present a brief census of the three main modeling approaches to galaxy evolution, namely hydrodynamical simulations, semi-analytic models, and semi-empirical models (SEMs). In Section \ref{sec|flavors} we focus on SEMs in their different flavors, i.e. interpretative, descriptive and hybrid, discussing the peculiarities and highlighting virtues and shortcomings for each of these variants. In Section \ref{sec|TOPSEM} we dissect a simple and recent hybrid SEM from our team to highlight some technical aspects. In Section \ref{sec|discussion} we offer some outlook on the prospective developments of SEMs. In Section \ref{sec|summary} we provide a short summary of this review.
\end{abstract}

\begin{BoxTypeA}{Key points}
\begin{itemize}
\item Semi empirical models (SEM) are effective, data-driven,
easily expandable and computationally low-cost approaches
to galaxy evolution, complementary to (semi-)analytic models and numerical simulations.
\item Descriptive SEMs produce realistic snapshots of the Universe at a given cosmic time. Interpretative SEMs serve to test specific hypotheses or consistency among different datasets. Hybrid SEMs feature elements partly in line with an interpretative and partly with a descriptive viewpoint.
\item SEMs are being exploited to address open issues in galaxy evolution like galaxy quenching mechanisms and the coevolution of galaxies and supermassive black holes. A strong synergy of SEMs with data science techniques could also be foreseen in the near future.
\end{itemize}
\end{BoxTypeA}

\section{Introduction: the triumvirate of galaxy formation models (SIMs, SAMs, and SEMs)}\label{sec|intro}

Acquiring a comprehensive picture for the formation and evolution of galaxies in a cosmological context is one of the main challenges in modern astrophysics (see \citealt{Mo2010,Silk2012,Maiolino2019,Cimatti2020}).
Whilst a well-defined standard model exists in particle physics, there is no equivalent in the realm of extragalactic astrophysics, where major unknowns still limit our understanding of galaxy formation and evolution. Leading theories suggest that galaxies form from the condensation and cooling of pristine gas within virialized dark matter (DM) structures called halos. Ironically, the evolution of such an `obscure' DM component is somewhat simpler and better understood than the `luminous' baryonic one, since allegedly the only force at play among DM particles is (Newtonian) gravity. Cosmological $N-$body simulations have been able to precisely delineate the formation and evolution of DM halos and to extensively characterize their statistical properties. In contrast, the physics regulating baryons is intrinsically complex, since it involves many physical processes occurring on vastly different spatial, time, and energy scales.\\

For example, it is well established that DM structures are formed in a hierarchical way, with larger halos gradually growing via the merging of smaller ones and/or mass accretion along filamentary structures of the cosmic web (e.g., \citealt{Springel2006,Shandarin2012, Vogelsberger2014, Libeskind2018,Martizzi2019,Wilding2021}; see also \citealt{Angulo2022}). However, observations suggest that star formation and stellar mass assembly in galaxies may follow an opposite, downsizing trend: massive galaxies form a significant part of their stellar mass earlier and rapidly, via a burst of intense star formation lasting less than a Gyr, while smaller ones form later on, over longer timescales of several Gyrs (see, e.g., \citealt{Cowie1996, Thomas2005,Thomas2010,MArtin2018,Merlin2019,Nanayakkara2022,Lah2023}). Explaining these contrasting trends is particularly challenging for current models, which tend to accurately predict the properties of local galaxies, but struggle somewhat in understanding the strong star formation episodes and large stellar masses observed at high redshift (see \citealt{Fontanot2007,Somerville2012, Somerville2015,Hirschmann2016,Bassini2020,Lovell2021,Hayward2021,Lustig2023,Dome2024}), unless specific assumptions are invoked such as a non-universal initial mass function (see \citealt{Baugh2005,Lacey2016,Fontanot2017,Lapi2024imf,Jeong2024}), reduced effects of dust attenuation (e.g., \citealt{Ferrara2023,Ferrara2024}), alternative cosmologies (e.g., \citealt{Menci2020,Menci2024}) or modified gravity theories (e.g., \citealt{McGaugh2024}).\\

Another example of an important and mysterious issue in galaxy evolution concerns not the birth of galaxies, but rather their \emph{deaths}. How do galaxies quench their star formation? Crucial hints for understanding star formation activity come from its relationship with galaxy morphology. In fact, up to very early epochs, on average it has been observed an intriguing correlation between star formation and morphology: elliptical galaxies or disk galaxies with a prominent bulge component are typically characterized by very low SFRs, while galaxies with a larger disk component tend to be star-forming (e.g., \citealt{Wuyts2011,Dimauro2022,Driver2022}). Nevertheless, as of now, there is no definitive observational evidence pinpointing the leading mechanism responsible for quenching galaxies: is it the presence of the stellar bulge itself that acts as the key driver behind quenching (e.g., \citealt{Cook2009,Martig2009,Gensior2020})? Or instead morphological evolution and star formation activity are indirectly connected by some other physical phenomenon able to quench the galaxy and at the same time shape its morphology? In this context, could quenching be caused by the energy/momentum feedback from the supermassive black hole residing at the centre of galactic bulges (e.g., \citealt{Silk1998,Fabian1999,Granato2004,Lapi2006,King2005,King2015rev,Grand2017,Lapi2018,Valentini2020,Goubert2024})? Or even just by the host halo mass surpassing a certain threshold and causing the retention of a hot atmosphere screening the galaxy from external gas inflows (e.g., \citealt{Birnboim2003,Keres2005,Dekel2006,Dekel2009,Dekel2023})? \\

In the last three decades, the numerous facets of such open, complex problems have been investigated mainly via three
ab-initio modeling approaches: hydrodynamical simulations (for a review, see \citealt{Naab2017}), semi-analytic models (for a review, see
\citealt{Somerville2015rev}), and analytic frameworks (for a
review, see \citealt{Matteucci2012}). \emph{Hydrodynamical simulations} (SIMs) allow to address the galaxy formation process in fine detail, as they can cope with the simultaneous evolution of DM, gas and stars. However, despite the recent increase in resolution and speed (mainly thanks to emulators and 'genetic' codes), many of the relevant physical processes still constitute sub-grid physics, while a detailed exploration of the parameter space is often limited by rather long computational times. Early SIMs struggled in forming cold, thin, and extended stellar disks, due to the overcooling of low-angular momentum baryons at high redshifts (see \citealt{Katz1991,Navarro2000,Abadi2003}). SIMs incorporating gas outflows
related to stellar feedback mitigated the issue by preventing
early cooling (see \citealt{Scannapieco2008,Governato2010,Brook2011}), and delaying the star formation
history toward the present (see \citealt{Brook2012,Stinson2013}), though at the price of yielding outcomes that are
sensitively dependent on the sub-grid recipes and the numerical
treatment of the feedback itself (e.g., \citealt{Scannapieco2012,Donnari2021,Habouzit2022,Crain2023}).
Subsequent developments have focused on reproducing the
overall galaxy structure, the metal abundance gradients, and the
scaling relations among the integrated properties of local disk
galaxies (see \citealt{Guedes2011,Aumer2013,Vogelsberger2014,Hopkins2014,Crain2015,Schaye2015,Colin2016,Ceverino2017,Pillepich2018,Hopkins2018,Dave2019}). Achieving these goals has required the introduction
of educated star formation thresholds, and the setting of the
feedback efficiency to high values. The latter has been
considered also in possible connection with the activity of a
central supermassive black hole (see \citealt{DeLucia2006,Grand2017,Valentini2020,Wellons2023}). The most recent SIMs are starting
to fully address the detailed spatial and kinematical structure of
galaxies and their cycle of multiphase gas across cosmic times (see \citealt{Grand2019,Pillepich2019,Vincenzo2019,Buck2020,Pandya2021,Feldmann2023}).\\

\emph{Semi-analytic models} (SAMs) are based on DM merger
trees extracted from $N-$body simulations (or generated via Monte Carlo procedures gauged on them), while the physics inside
dark halos is modeled via parametric expressions set on
(mainly) local observables. These models are less computationally
expensive than hydro simulations and enable to more
clearly disentangle the relative role of the diverse physical processes, though the considerable number of fudge parameters describing the underlying physics can lead to degenerate solutions and restrict somewhat the predictive power, especially toward high redshift. Early attempts based on simple recipes
for cooling and stellar feedback from supernova
explosions yielded encouraging results in reproducing the
properties of local disk galaxies such as sizes, scaling relations,
and statistics (see \citealt{Kauffmann1993,Lacey1993,Cole2000,Baugh2005}). Through the years, these
basic prescriptions have been progressively refined to describe
additional processes and further improve the agreement with
observations, even toward high redshifts. Specifically, modern
SAMs incorporate energy feedback from
accreting supermassive black holes (see \citealt{Croton2006,Monaco2007,Somerville2008,Benson2012,Fontanot2020}), merger-driven bursts of
star formation and dust absorption/emission effects (see \citealt{Baugh2005,Cook2009,Lacey2016}), recycling of
blown-out gas via galactic fountains (see \citealt{Herniques2015,Croton2016}), metal enrichment in gas and stars (see
\citealt{Cousin2016,Hirschmann2016}), multiphase
treatment of atomic and molecular gas components (see
\citealt{Somerville2015,Lagos2018,Baugh2019}),
radial structure and gradients (see \citealt{Stevens2016,Henriques2020}) and related transport processes of gas
and stars (see \citealt{Forbes2019}), and the effects of DM halo growth rate, gas fraction and angular momentum increase/dissipation on the emergence of galaxy properties (see \citealt{Cook2009,Cattaneo2017,Barrera2023,Mo2024}).\\

It is also worth mentioning the role of frameworks that admit fully analytic solutions. They are necessarily
based on approximate and spatially/time-averaged descriptions
of the most relevant astrophysical processes; however, their
transparent, handy, and predictive character often pays off on
some specific issues. Pioneering works were focused on the
chemical evolution of the Milky Way, and highlighted the
relevance of gas inflow and outflow processes in reproducing
the metal abundance of the solar neighborhood (see \citealt{
Schmidt1963,Talbot1971,Tinsley1974,Pagel1975,Hartwick1976,Chiosi1980,Matteucci1986,Edmunds1990}). Successive developments
concerned the mechanisms leading to dust production (see \citealt{Dwek1998,Hirashita2000,Inoue2003,Zhukovska2008}), the abundance gradients in the Galactic disk (see
\citealt{Chiappini2001,Naab2006,Grisoni2018}), steady-state equilibrium models among star formation,
inflows and outflows (see \citealt{Bouche2010,Dave2012,Lilly2013,Pipino2014,Feldmann2015,Pantoni2019,Lapi2020}), effects of
the initial mass function and of stellar yield models (see  \citealt{Molla2015,Recchi2015}), differential and/or
selective winds (see \citealt{Recchi2008}), dichotomy among
active and passive galaxies (see \citealt{Spitoni2017}), and detailed
inside-out growth of galaxy disks with radial mixing (see
\citealt{Andrews2017,Frankel2019}).\\

All in all, it is fair to state that in relation to the aforementioned open problems about downsizing, quenching and galaxy morphological evolution, all these ab-initio theoretical approaches somewhat struggle in highlighting the primary physical mechanism responsible for them. They offer insights into the influence of the diverse physical processes, yet their outcomes are contingent on the chosen physical models, sub-grid recipes and resolution (see \citealt{Scannapieco2012,Donnari2021,Crain2023}). To partly bypass
these shortcomings, a  complementary approach to probe galaxy evolution has emerged, namely the so-called \emph{semi-empirical
models} (SEMs) as pioneered by different research groups  (e.g., \citealt{Hopkins2006,Hopkins2009,Moster2013,Moster2018,Behroozi2013,Behroozi2019,Behroozi2020,Shankar2014,Mancuso2016b,Mancuso2016a,Mancuso2017,Buchan2016,Grylls2019,Hearin2022,Drakos2022,Fu2022,Fu2024,Boco2023,Zhang2023,Zhang2024}). SEMs adopt an ‘effective’ approach to galaxy evolution: they do not attempt to model the small-scale physics regulating the baryon cycle from first principles, but marginalize over it by exploiting empirical relations between the spatially-averaged properties of galaxies (e.g., stellar mass, SFR, specific SFR) and DM halos (e.g., halo mass, accretion rate, or circular velocity), derived from their relative abundance or from analytic parameterizations. The value of these models stands in that they feature by design a minimal set of assumptions and parameters gauged on observations. By focusing just a few aspects they allow to transparently investigate specific and still unclear/debated aspects of galaxy formation and evolution, in a bottom up fashion. Moreover, by empirically linking different observables, they can test for possible inconsistencies among distinct datasets, which can often occur given the significant observational systematics. Finally, these frameworks are particularly helpful when galaxy formation recipes must be coupled with those from other branches of astrophysics (e.g. stellar evolution, planetary science, radiative transfer codes, etc.), and therefore it is worth to
minimize the uncertainties/hypotheses at least on the former side by exploiting basic data-driven inputs (e.g., \citealt{Lapi2024,Roy2024}). 
However, semi-empirical models are not free of downsides. By construction they strongly rely on data and, therefore, it is essential to input them with robust determinations of galaxy statistics and scaling relations between baryonic and halo properties to build a successful framework. \\

\section{SEMs in their different flavors: descriptive, interpretative, hybrid}\label{sec|flavors}

The basic setup of a SEM requires to establish a connection between the properties of DM halos and galaxies. Specifically, starting from a DM halo statistics and/or catalog extracted from a $N-$body simulation, SEMs populate the halos with galaxies via matching the statistical distributions of two different physical quantities, one related to DM and the other to the baryonic component. Such an `abundance matching' technique implicitly assumes the existence of a monotonic relationship between two quantities, $X$ and $Y$, one related to halos and one to baryons, and matches their cumulative statistical distributions $P(>X)$ and $P(>Y)$. For example, on the DM side $X$ can be the halo mass $M_{\rm H}$ and $P(>M_{\rm H})$ can be easily constructed from the halo mass function $N(M_{\rm H},z)$ extracted from $N-$body simulations. On the baryonic side $Y$ can be the stellar mass $M_\star$ and $P(>M_\star)$ can be built from the observed galaxy stellar mass function $N(M_\star,z)$. The main problem of such an abundance matching technique is that perfect monotonic relations do not 
always efficiently capture all the physics involved: galaxies and halos will often feature a certain distribution in the $X-Y$ plane, spread out by the degeneracies with many other nuisance parameters. However, abundance matching techniques do not allow to derive such a full distribution, but just an average relation $Y(X)$ and, even when scatter is kept into account, it is often assumed Gaussian and guessed a priori (see \citealt{Aversa2015}). For this reason, every SEM based on abundance matching requires an assumption of monotonicity between two quantities and a free parameter, namely the scatter on their relation. Nevertheless, abundance matching constitutes a powerful tool to understand how different observables relate to each other, and to address systematic uncertainties in the statistical distributions from which the cumulative probabilities are built (especially the baryonic related ones).\\

\begin{figure}[t]
\centering
\includegraphics[width=.49\textwidth]{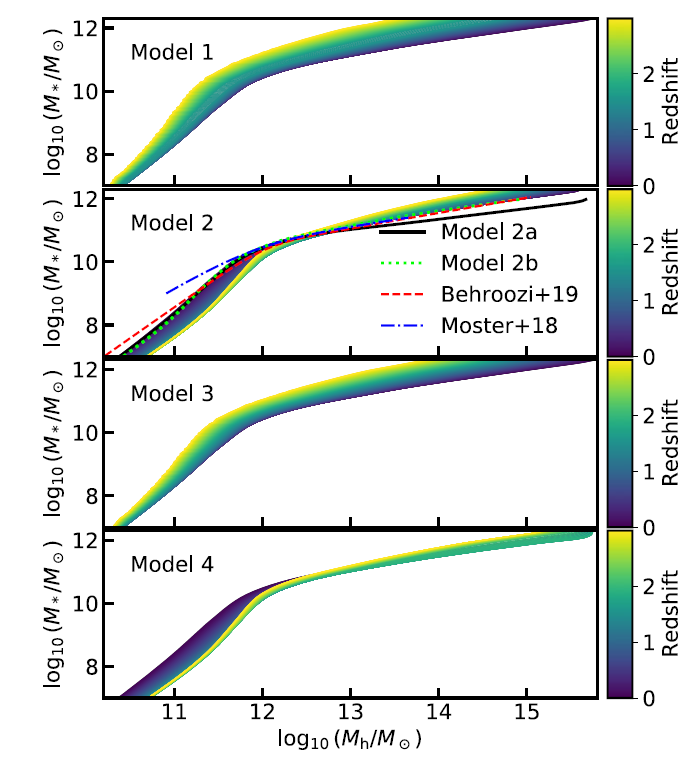}~~~~~~~~~
\includegraphics[width=.4\textwidth]{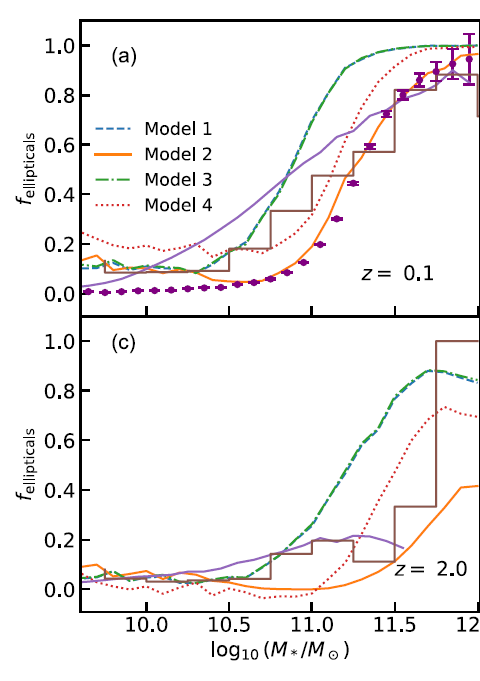}
\caption{Figures from \cite{Fu2022} highlighting some results of the \texttt{DECODE} model. The left panel illustrates the SMHM relationship at different redshift (color-coded) computed via abundance matching of the halo mass and stellar mass functions. The four panel refer to different models for the shape and evolution of the stellar mass function; in the second from the top, the local relations from the \texttt{EMERGE} (dot-dashed line) and \texttt{UNIVERSEMACHINE} (dotted line) SEMs are reported. The right panel illustrates the fraction of elliptical galaxies vs. the stellar mass at two representative redshifts $z\approx 0.1$ (top panel) and $z\approx 2$ (bottom panel), for the same four models. Data from the SDSS survey (magenta circles), the \texttt{GALICS} SAM (solid magenta line; see \citealt{Koutsouridou2022}) and the \texttt{TNG100} SIM (magenta histogram; see \citealt{Nelson2019}) are reported.}\label{fig|decode}
\end{figure}

The very first attempts to link galaxies and host halos via abundance matching involved the luminosity and halo mass (e.g., \citealt{Vale2004,Vale2006,Shankar2006}). Through the years, the most commonly employed relation has been the one between the halo mass $M_{\rm H}$ and stellar masses $M_\star$, called stellar mass-halo mass (SMHM) relation (e.g., \citealt{Hopkins2009,Conroy2009,Behroozi2010,Moster2010,Yang2012,Zavala2012,Behroozi2013,Moster2013,Shankar2014,RodriguezPuebla2017,Tollet2017,Grylls2019,Fu2022}). More recently, people have started using more sophisticated, additional mappings between the time derivative of the masses (e.g., \citealt{Fu2024}), i.e. the star formation rate $\dot M_\star$ vs. the halo accretion rate $\dot M_{\rm H}$, or even between the specific rates (see \citealt{Boco2023}), i.e. the specific SFR $\dot M_\star/M_\star$ and the specific halo accretion rate $\dot M_{\rm H}/M_{\rm H}$. A recent example of a SEM following the aforementioned  foundational framework is constituted by the \texttt{DECODE} model (see \citealt{Fu2022}), where abundance matching techniques are exploited to derive the SMHM relation and the merger histories of central and satellite galaxies. Specifically, the main focus is to highlight the dependence of these derived quantities on the observational determination of the stellar mass function up to the cosmic noon at redshift $z\sim 2$. A few outcomes of \texttt{DECODE} can be found in Figure \ref{fig|decode}; these showcase that when different renditions of the local stellar mass function and of its redshift evolution are adopted, the SMHM relation and the fraction of elliptical galaxies change accordingly (at fixed theoretical assumptions, e.g. role of major mergers in their formation path). Remarkably, it turns out that only SMHM relations derived from stellar mass functions featuring large number densities of massive galaxies and significant redshift evolution (`Model 2' in the Figure) can simultaneously reproduce the local abundances of satellite galaxies, the major merger statistics as inferred from galaxy pairs, and the local fraction of elliptical galaxies. This is an example of how a SEM can be exploited to have a handle on possible systematic uncertainties in the observational determination of a statistical distribution like the stellar mass function. For this reason, SEMs with this flavor may be referred to as \emph{interpretative}.\\

\begin{figure}[t]
\centering
\includegraphics[width=.8\textwidth]{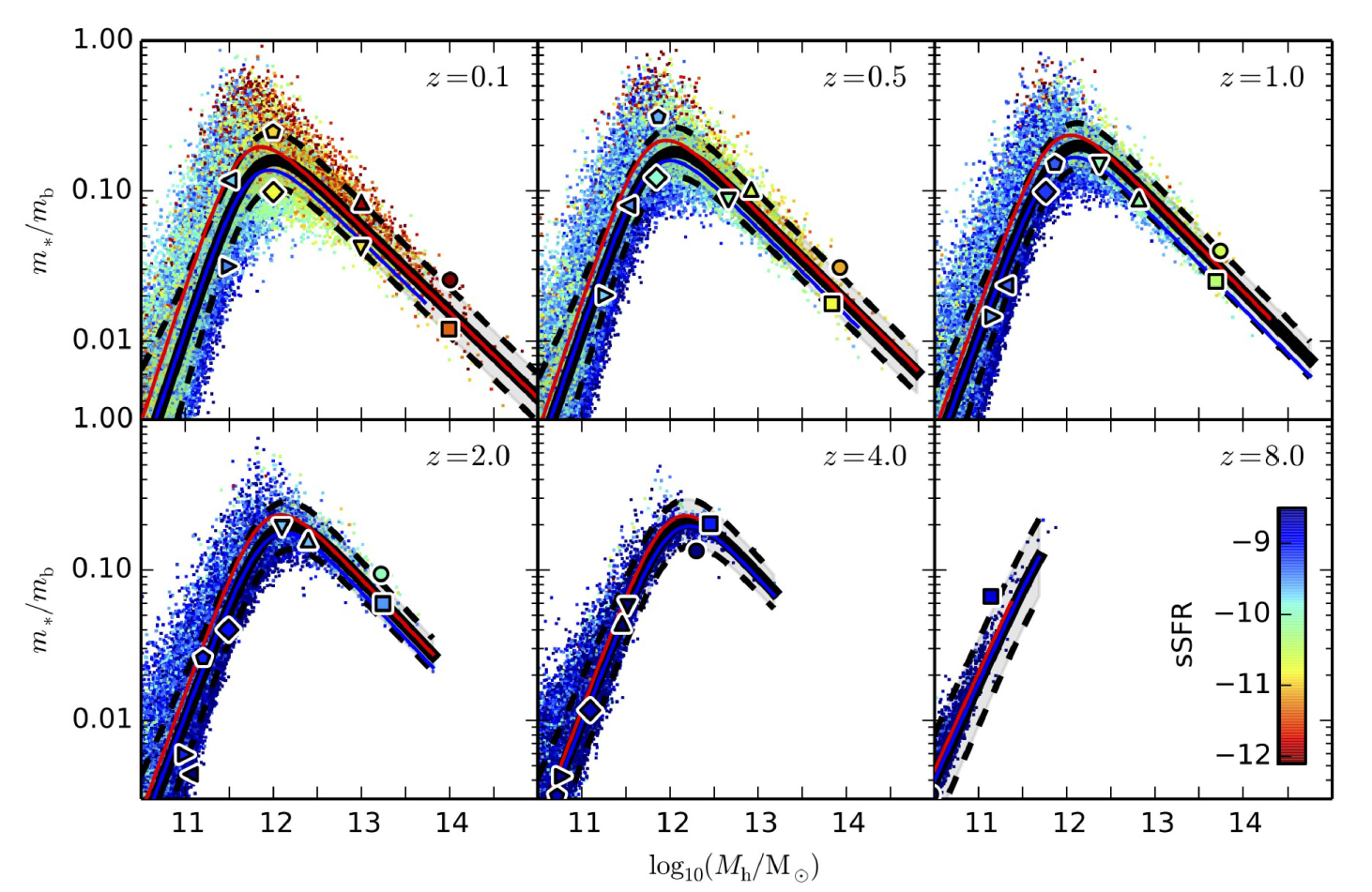}
\caption{Figure from \cite{Moster2018} illustrating the redshift-dependent SMHM derived from \texttt{EMERGE} at different redshifts from $z\approx 0.1$ to $z\approx 8$ (from top left to bottom right), expressed in terms of the integrated conversion efficiency $M_\star/0.16\, M_{\rm H}$ vs. the halo mass $M_{\rm H}$. Color-code refers to the logarithm of the specific SFR $\log \dot M_\star/M_\star$ in units of yr$^{-1}$. Solid black lines show the median and the dashed black lines show the $1\sigma$ scatter. The red and blue solid lines are the median for quenched and starforming galaxies. The large symbols in each panel represent eight individual systems that have been selected
from the upper and lower $1\sigma$ contours for four halo masses at $z\approx 0.1$, and traced back in cosmic time.}\label{fig|emerge}
\end{figure}

\begin{figure}[t]
\centering
\includegraphics[width=.7\textwidth]{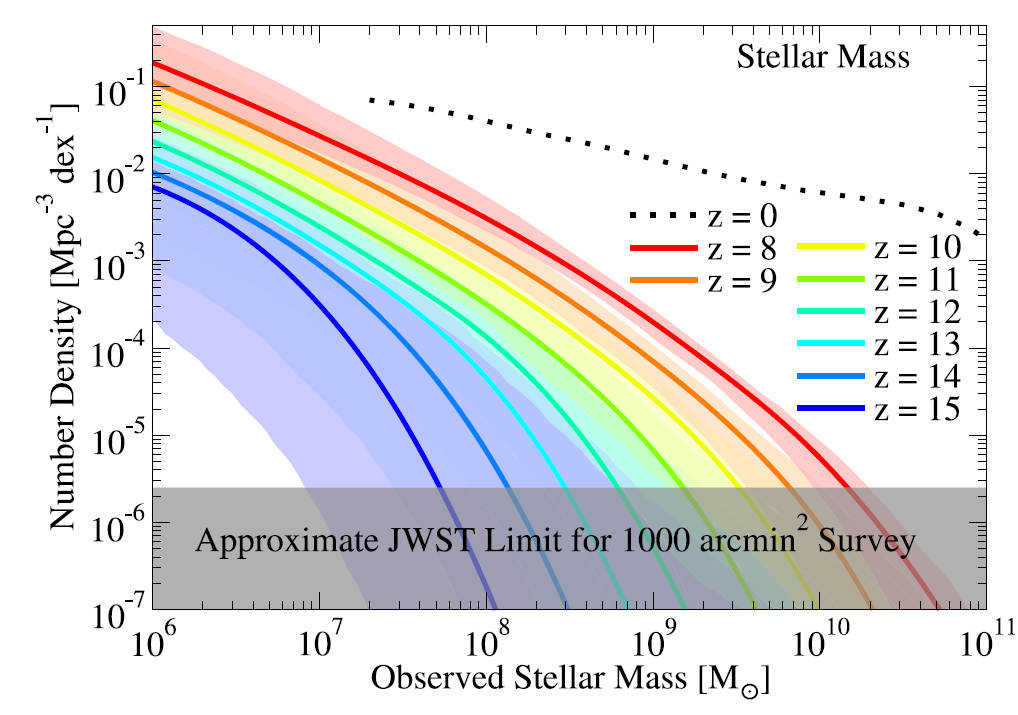}
\caption{Figure from \cite{Behroozi2020} highlighting predictions of \texttt{UNIVERSEMACHINE} for the stellar mass function at high redshift from $z\approx 8$ to $z\approx 15$ (color-coded), in terms of median values (solid lines) and $1\sigma$ confidence intervals (colored areas). The local stellar mass function is also reported as a dotted black line. The approximate limit of a JWST survey over $1000$ arcmin$^2$ is displayed as a grey shaded area.}\label{fig|universemachine}
\end{figure}

An alternative SEM framework envisages to connect DM halo and galaxy properties by assuming a parametric form of an empirical relation expected between two quantities (inspired from observational evidence or theoretical arguments), and by fitting such parameters by comparison with other independent observables. An example could be to parameterize the local SMHM relation in terms of a double powerlaw behavior $M_\star(M_{\rm H},z) = \epsilon\, M_{\rm H}/[(M_{\rm H}/M_c)^{\alpha}+(M_{\rm H}/M_c)^{\beta}]$, with $\epsilon$ a normalization, $M_c$ a characteristic mass, $\alpha$ and $\beta$ two slopes at large and small halo masses. Furthermore, the redshift evolution of the relation may in turn be rendered via a polynomial dependence of the above quantities in terms of the scale factor $a\equiv 1/(1+z)$, e.g. $\epsilon=\epsilon_0+\epsilon_z\,(1-a)$ with $\epsilon_0$ and $\epsilon_z$ the fitting parameters. Plainly, the complexity of such a description depends on the number of parameters and quantities designed to be fitted (e.g., \citealt{Moster2018,Behroozi2019,Grylls2020}).
A specific example of this approach is constituted by \texttt{EMERGE} (Empirical ModEl for the formation of GalaxiEs; see \citealt{Moster2018}). It assigns galaxies to DM halos by parameterizing a redshift-dependent SFR vs. halo accretion rate relation, and then fit for the parameters exploiting the observational constraints on stellar mass function, cosmic SFR density and quenched galaxy fractions. The added value of this approach is that the SMHM relation and its scatter, together with the associated redshift evolution, comes out as predictions of the model (see Figure \ref{fig|emerge}). Another example worth mentioning is \texttt{UNIVERSEMACHINE} (see \citealt{Behroozi2019}). This SEM assumes a parameterized relation between the SFR and the peak circular velocity (related to the maximum mass attained over the halo’s assembly history) which, along with quenched galaxy fractions, is used to assign a SFR probability distribution to each DM halo. These distributions are exploited to compute individual SFRs for halos and then the latter quantity is integrated in time to derive galaxy stellar mass growth tracks. The galaxy stellar mass function, cosmic SFR, correlation functions and other quantities are predicted as outputs of the model. In a subsequent, very interesting development (see \citealt{Behroozi2020}), \texttt{UNIVERSEMACHINE} has been also exploited to generate mock galaxy catalogs and light-cones over an extended redshift range up to $z\sim 15$, providing specific predictions for JWST surveys to be confronted with recent and upcoming data (see Figure \ref{fig|universemachine}). SEMs like the two above are designed with many parameters to be fitted, and are closer in this respect to a full-fledged SAM. However, they still keep a close link between parameters and observations typical of the SEM spirit, and allow to  choose an educated, empirical parameterization not necessarily anchored to an ab-initio theoretical modeling of the physical processes at play. This approach has the added value of easing the removal of degeneracies among parameters, the exploration of the full parameter space, and the optimization of parameters via standard model selection statistics. SEMs with this flavor may be referred to as \emph{descriptive}.\\

Finally, other frameworks try to complement the basic setup of a descriptive SEM on some galaxy properties, with the integration of interpretative elements to constrain some underlying physical processes. Like any other SEMs, such approaches are expressly designed to rely on a limited set of assumptions and parameters, rendering the model almost fully data-driven. But then specific physical hypotheses are applied on top of this basic scheme and each of the putative triggers for a phenomenon is systematically tested, in isolation or in combination with others, against the observed abundances and properties of galaxies at different cosmic epochs. Such an approach is different from a descriptive SEM, because it goes beyond establishing relationships between various properties of halos and galaxies. Instead, it aims to elucidate how different theoretical hypotheses on a phenomenon may influence these relationships. The effect of adding, removing, or changing such a physical hypothesis for a phenomenon will manifest visibly in the outputs and can be directly compared to observational data. Such kind of SEMs may be helpful to address the longstanding issues in galaxy evolution recalled in Section \ref{sec|intro}. For example, the effectiveness of such novel approaches has been recently demonstrated in the \texttt{TOPSEM} model (TwO Parameter Semi-Empirical Model; see \citealt{Boco2023}), which is capable of reproducing the galaxy morphology with a minimal set of assumptions and parameters. Specifically, after constructing a mock catalog of galaxies via abundance matching, \texttt{TOPSEM} tests a theoretical hypothesis for the morphological evolution of galaxies, strictly linking it to the host DM halo growth (see next Section for details). Physical scenarios for quenching and downsizing are also starting to be investigated with this kind of approach (e.g., \citealt{Fu2025}; see Section \ref{sec|discussion}). All in all, such a way of testing physical hypotheses on top of empirical frameworks embodies partly interpretative and partly descriptive components, and for this reason this type of SEM may be referred to as \emph{hybrid}. \\

\section{Dissecting a hybrid SEM}\label{sec|TOPSEM}

In this Section we present some details of a recent model from our research team, dubbed \texttt{TOPSEM}: Two-Parameters Semi Empirical Model (see \citealt{Boco2023}), as an example of the very simple and effective approach typical of a SEM. We will show that on the one hand this straightforward model is able to reproduce crucial aspects of galaxy evolution, such as the stellar mass assembly in galaxies, and on the other hand it can be exploited to test specific hypotheses on the origin of galaxy morphology, so addressing some of the pressing issues recalled in Section \ref{sec|intro}. \texttt{TOPSEM} relies on one main assumption, one boundary condition, and two parameters associated to them. The assumption is about a monotonic correlation between galaxies' specific star formation rate (sSFR) and specific halo accretion rate (sHAR), with sSFR being the ratio between star formation rate and stellar mass of a galaxy and sHAR being the ratio between halo accretion rate and halo mass. This assumption means that galaxies with larger sSFR reside in halos with larger sHAR. The model relies on abundance matching of these specific rates to derive the sSFR-sHAR relationship. Building on that, DM halos are populated with galaxies of given SFR and stellar masses, effectively constructing a mock catalog of galaxies equipped with their own stellar mass assembly history.\\

Such a mock catalog is used to test a theoretical hypothesis on the morphological evolution of galaxies, in the spirit of the hybrid SEMs described in the previous Section. This hypothesis is based on a two-phase galaxy evolution scenario, where the formation of bulges and disks is linked to the two-mode DM halo accretion histories. Specifically, $N-$body simulations analyses (e.g., \citealt{Wechsler2002,Zhao2003,Zhao2009,Diemand2007,Hoffman2007,Ascasibar2008,More2015,Hearin2021}) have highlighted that mass accretion history of halos can be roughly divided into two distinct phases: an early fast accretion phase dominated by major mergers and violent collapse, which shapes the structure of the inner halo potential well, and a later slow accretion phase characterized by a smoother DM accretion or minor mergers and usually dominated by `pseudo-evolution' (\citealt{Diemand2007,Diemer2013,Zemp2014,More2015}), which contributes to growing the halo outskirts though not altering the central structure. The hypothesis tested is that bulges and spheroidal components are formed during the early fast accretion phase, when violent accretion processes and dynamical friction between giant gas clumps leads to a quick loss of angular momentum also for baryonic matter that rapidly sinks to the very central region and triggers violent star formation episodes (e.g., \citealt{Lapi2018,Pantoni2019}). Galactic stellar disks are instead formed during the later slow accretion phase when baryons can retain part of their angular momentum and are not directly funneled toward the center. Such an idea builds upon the pioneering works of \cite{Mo2004} and \cite{Cook2009} and can qualitatively explain stellar archaeological results showing stellar population of bulges/spheroids to be older, $\alpha-$enhanced and rapidly generated, and stars in disks to be younger and formed over longer timescales (see, e.g., \citealt{Cowie1996,Chiappini1997,Thomas2005,Thomas2010,Gallazzi2006,Johansson2012,Corteau2014,Pezzulli2016,Grisoni2017,Bellstedt2024,Mo2024}). In \texttt{TOPSEM} the interplay between the transition time from fast to slow accretion and the quenching time would lead to the generation of different types of galaxies, from grand-design disks to pure ellipticals. As shown in section \ref{sec|morphology}, the model naturally predicts the stellar mass function for bulges and ellipticals and the fraction of ellipticals, that are found to be in agreement with the observational determination from the GAMA survey by \cite{Moffett2016b,Moffett2016a}.\\

\subsection{Empirical relations for halos and galaxies}

\begin{figure}[t]
\centering
\includegraphics[width=.49\textwidth]{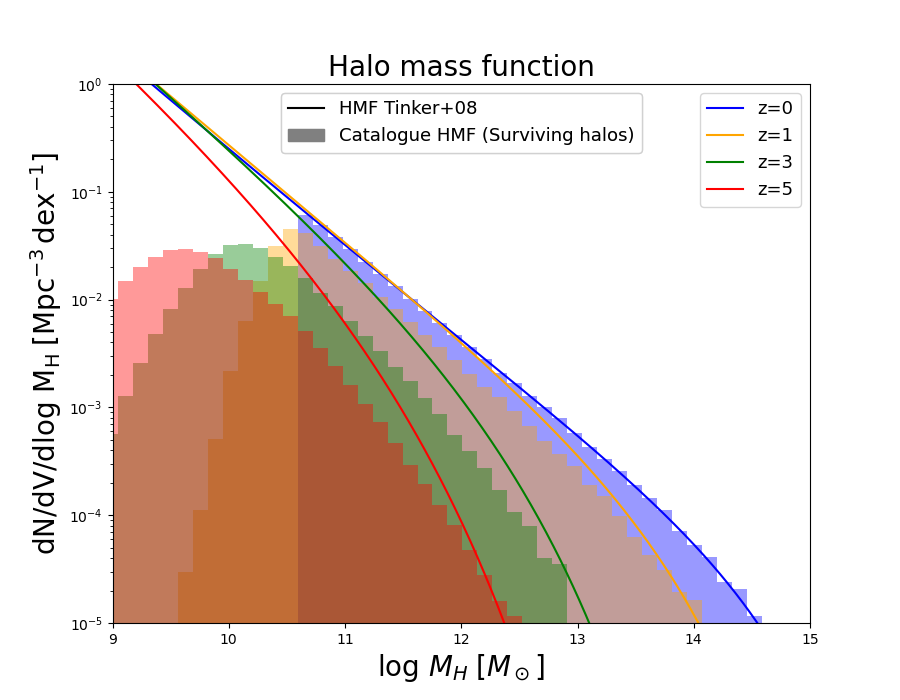}~~~~~~~~~
\includegraphics[width=.49\textwidth]{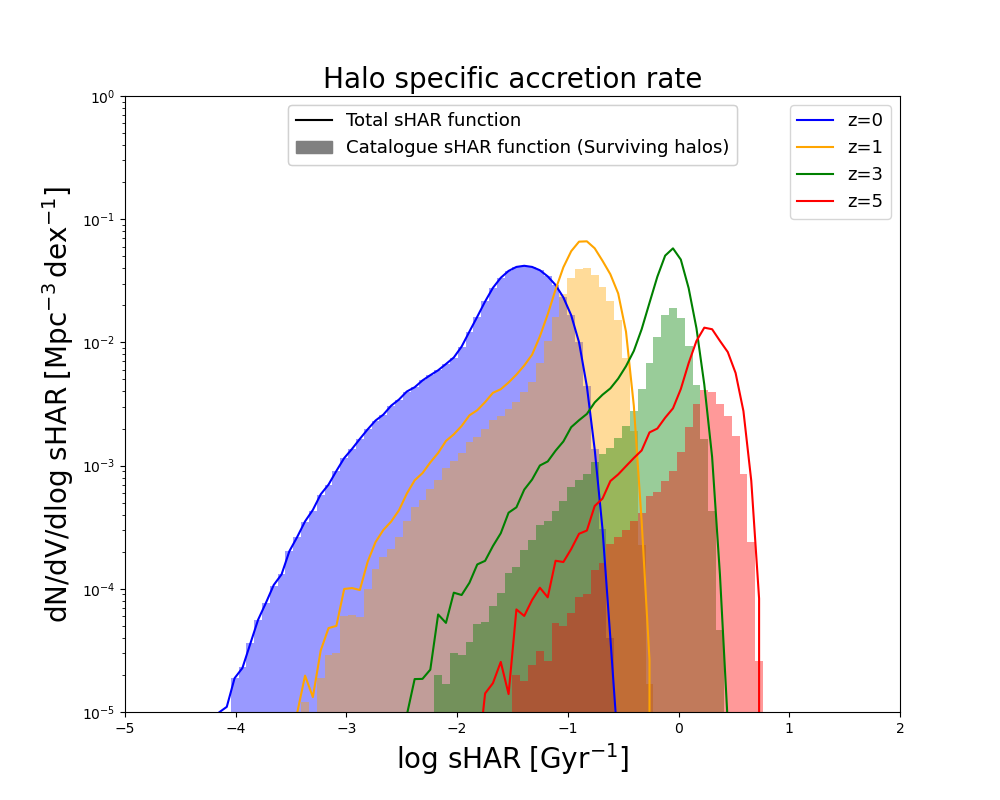}
\caption{Figure from \cite{Boco2023} showing the halo mass function (left panel) and sHAR function (right panel) at different redshifts: blue $z = 0$, orange $z = 1$, green $z = 3$, and red $z = 5$. Histograms are the distributions for a mock catalog of surviving halos, whereas solid lines refer to the reconstructed total distribution for all the halos.}
\label{fig|distributions}
\end{figure}

The first step for every SEM is to build a catalog of DM halos, which is then populated with galaxies by means of empirical relations. Halo statistics and their evolution are completely set by $N-$body simulation results. In \texttt{TOPSEM} a discrete halo catalog at $z=0$ is built by sampling the halo mass function  determination from \cite{Tinker2008}, tracing the halo growth history backward in time using the \texttt{DIFFMAH} code from \cite{Hearin2021}. The catalog constructed with this method naturally comprises and tracks the evolution of all the halos surviving at $z=0$, but does not include the non-surviving halos, i.e., the ones which merge or are disrupted at higher redshift. The catalog is used to extract statistical distributions of meaningful halo properties, such as their mass and specific halo accretion rate, which are useful for building the model. Figure \ref{fig|distributions} shows the redshift evolution of the halo mass function (left panel) and of the sHAR function (right panel). Histograms refer to the halos in the catalog, while solid lines to the reconstructed distribution for all the halos at a given redshift, both surviving and non-surviving. In order to reconstruct the non-surviving halo distribution, it is assumed that each non-surviving halo of a given mass follows the same sHAR distribution as surviving halos of the same mass. This assumption is supported by the fact that, prior to becoming satellites of larger halos, non-surviving halos can be treated as centrals, exhibiting a sHAR distribution statistically similar to that of other halos with the same mass (see \citealt{Boco2023} for details). In this way, the effect of non-surviving halos is accounted for in TOPSEM results. As expected, higher redshift halos tend to have smaller masses and higher specific accretion rates. \\

\begin{figure}[t]
\centering
\includegraphics[width=.5\textwidth]{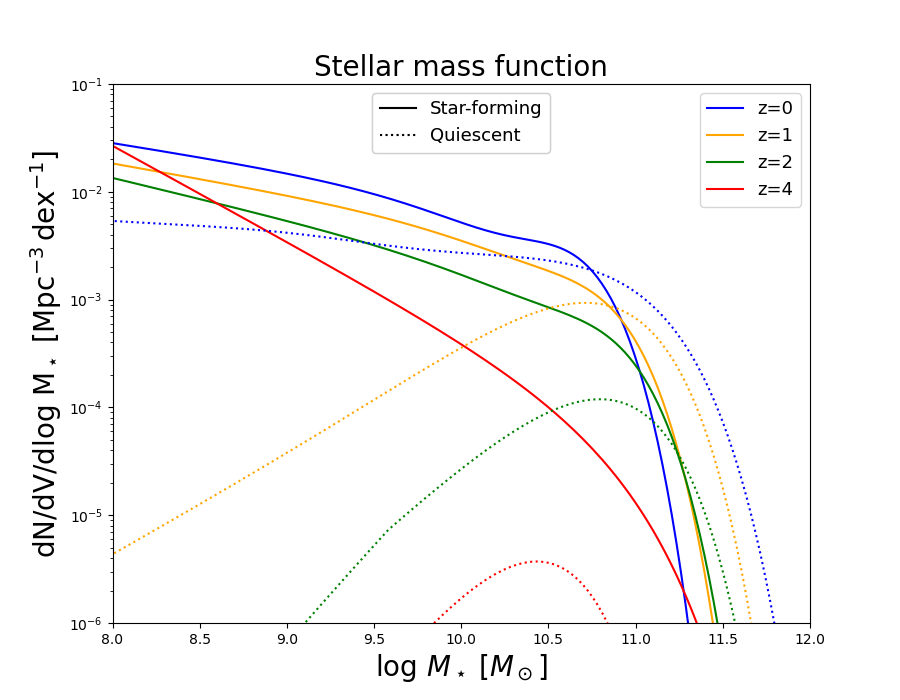}~~~~~~~~~
\includegraphics[width=.5\textwidth]{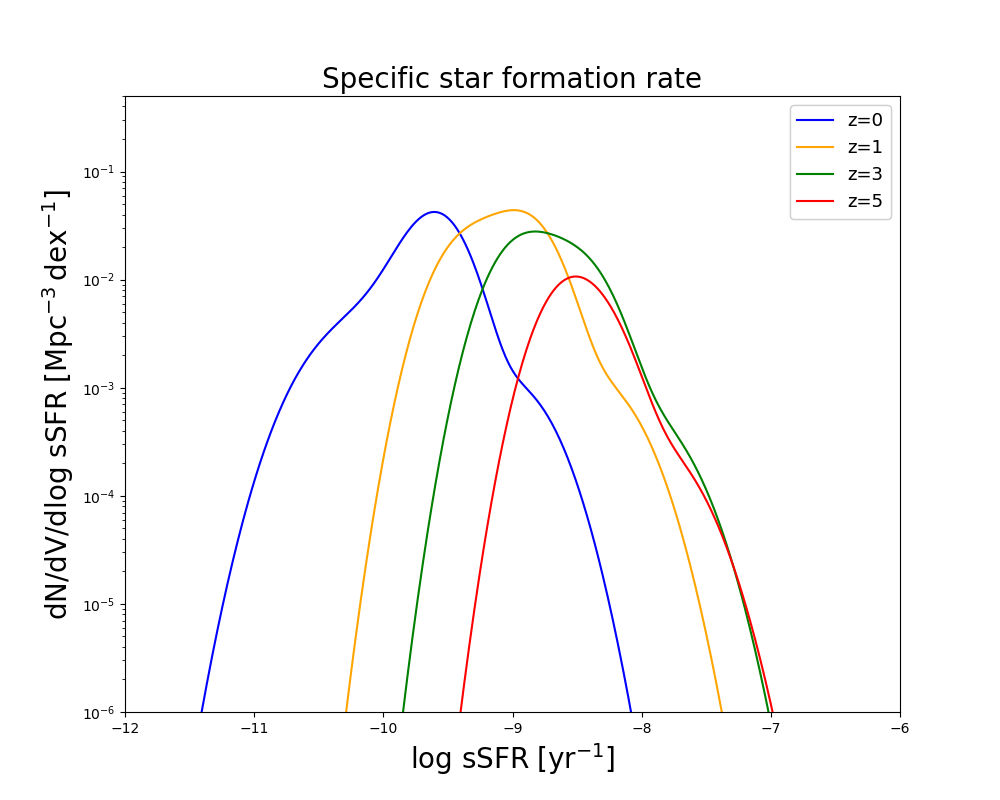}
\caption{Figure from \cite{Boco2023} showing the stellar mass function (left panel) from \cite{Davidzon2017} for star-forming (solid lines) and quiescent galaxies (dotted lines), and the sSFR distribution (right panel) from the convolution of the stellar mass function of active galaxies and the main sequence. Colors are as in Figure \ref{fig|distributions}.}
\label{fig|stellar distributions}
\end{figure}

For the treatment of the complex baryonic component, \texttt{TOPSEM} relies on empirical quantities and relations at different redshifts, which provide a snapshot of galaxy properties at any cosmic time. These include the stellar mass function both for starforming and quiescent galaxies (e.g., \citealt{Ilbert2013,Tomczak2014,Bernardi2017,Davidzon2017,Weaver2023}), describing how stellar mass is distributed among galaxies, and the main sequence of starforming galaxies (e.g., \citealt{Daddi2007,Rodighiero2011,Sargent2012,Bethermin2012,Rodighiero2015,Speagle2014,Whitaker2014,Ilbert2015,Schreiber2015,Mancuso2016b,Dunlop2017,Bisigello2018,Pantoni2019,Lapi2020,Popesso2023,Huang2023}), a well-known correlation between stellar mass and SFR at given redshift. In particular, \texttt{TOPSEM} adopts the fits proposed by \cite{Davidzon2017}
for the stellar mass function (left panel in Figure \ref{fig|stellar distributions}) and the main sequence determination by \cite{Popesso2023}. By convolving stellar mass function and main sequence (see \citealt{Boco2021a,Boco2023}), \texttt{TOPSEM} derives the sSFR distributions competing to star-forming galaxies at any given cosmic epoch (right panel in Figure \ref{fig|stellar distributions}).\\

\subsection{Populating halos with galaxies}

In the previous Section we have described the main ingredients needed to construct a SEM: the results of $N-$body simulations for the halo accretion history, and empirical quantities such as the stellar mass function and the main sequence for the baryonic component. The aim of a SEM is to connect these two components in an empirical way. \texttt{TOPSEM} establishes this link via abundance matching between sSFR and sHAR, so innovating with respect to the standard abundance matching between stellar mass and halo mass. In fact, stellar and halo masses can depend on the history of mass accretion, while specific quantities better capture the situation at a given time. This is because the SFR of a galaxy is expected to be somewhat linked to the halo accretion rate, since a faster DM accretion would correspond to a faster gas inflow and, consequently, to more gas available to form stars. On top of that, star formation efficiency, the amount of baryons converted into stars, can heavily change for objects with different masses (e.g., \citealt{Moster2010,Moster2013,Moster2018,Behroozi2013,Behroozi2019,Aversa2015,RodriguezPuebla2017}), possibly creating some scatter around the mean SFR-HAR relation. 
To marginalize over this effect, \texttt{TOPSEM} performs abundance matching between two mixed quantities sSFR $=\psi/M_\star$, and sHAR $=\dot M_H/M_H$, searching for a relation $\psi/M_\star=f(\dot M_H/M_H)$ where $f(\cdot)$ is a monotonic function to be found. Therefore, the main assumption of the model is the monotonicity in the sSFR-sHAR relation and the first free parameter is the scatter $\sigma_{\log sSFR}$ around this relation. Before giving the details of the abundance matching procedure and the results, two important remarks are in order: 

\begin{itemize}

\item sSFR alone is not useful to fully characterize a galaxy star formation history if the stellar mass at a certain time is not known (translating sSFR to SFR requires the knowledge of $M_\star$). For this reason \texttt{TOPSEM} needs a boundary condition for the stellar masses of the mock galaxies. These are initialized at $z\approx 0$, since we dispose of more complete and robust observational data in the local Universe with respect to high redshift. Such initialization is done on the basis of the $z\approx 0$ SMHM relation (see \citealt{Reyes2012,Mandelbaum2016,Lapi2018}), and 
the related scatter $\sigma_{\log M_\star}$ is the second parameter of the model. Then star formation histories of galaxies are followed backwards in time. Note that \texttt{TOPSEM} uses only the $z\approx 0$ SMHM as boundary condition, but the evolution of the relation is a prediction not an input.

\item \texttt{TOPSEM} assumes that only star-forming galaxies follow the sSFR-sHAR relation, not quenched ones. Indeed, the sSFR function displayed in Figure \ref{fig|stellar distributions} is valid only for star-forming galaxies; sSFR for quiescent ones are highly uncertain and rather disperse. Moreover, if the relation  between sSFR and sHAR is applied to all the halos and galaxies, it would naturally provide an explanation for quenching related to the accretion history of the DM halo: quenched galaxies, having low sSFR, would be classified as the ones embedded in low sHAR halos. However, the physical reasons of quenching are still largely debated and possibly related to complex baryon physics. Since the aim of \texttt{TOPSEM} is not to pin down the physical mechanism leading to quenching, it tries to be as agnostic as possible in regards to this problem. Therefore the monotonic relation between sSFR and sHAR is assumed to be valid only up to the moment of quenching; thereafter, stellar mass does not grow any longer, independently of the accretion history of the DM halo. Despite this, \texttt{TOPSEM} do keep track of passive galaxies in a fully empirical way, by following the evolution of the quiescent galaxy stellar mass function across various redshifts. In essence, the quiescent galaxy distribution describes the number of quiescent galaxies with a given stellar mass at different redshifts, and its temporal evolution provides information on the number of galaxies quenching at a specific redshift. By reproducing this distribution, TOPSEM is able to infer the number of quiescent galaxies across cosmic time and their redshift of quenching $z_Q$ (for details see \citealt{Boco2023}).\\
\end{itemize}

In \texttt{TOPSEM} the sSFR vs. sHAR relation is derived via abundance matching without any a-priori parametrization, according to:
\begin{equation}
\int_{\log\rm sSFR}^{\infty}\,{\rm d}\log {\rm sSFR}'\,\frac{{\rm d}p}{{\rm d}\log {\rm sSFR}'}=\int_{-\infty}^{\infty}\,{\rm d}\log {\rm sHAR}'\,\frac{{\rm d}p}{{\rm d}\log {\rm sHAR}'}\times\frac{1}{2}\,{\rm erfc} \left[\frac{\log(\rm {\rm sHAR(sSFR,z)/sHAR'})}{\sqrt{2}\,\sigma_{\log\rm sSFR}}\right],
\label{eq:abma_s_eq}
\end{equation}
where ${\rm d}p/{\rm d}\log {\rm sHAR}$ and ${\rm d}p/{\rm d}\log {\rm sSFR}$ are the distributions in sHAR and sSFR. The sSFR(sHAR,$z$) relation is derived by numerically solving Equation (\ref{eq:abma_s_eq}). This corresponds to finding the most suitable sSFR value for each sHAR, in order to equate the two integrals, thus matching the cumulative distributions of sSFR and sHAR. This is the essence of the abundance matching procedure. The quantity $\sigma_{\log\rm sSFR}$ appearing at the denominator in the right-hand term represents the scatter on the relation.

\begin{figure*}
\centering
\includegraphics[scale=0.45]{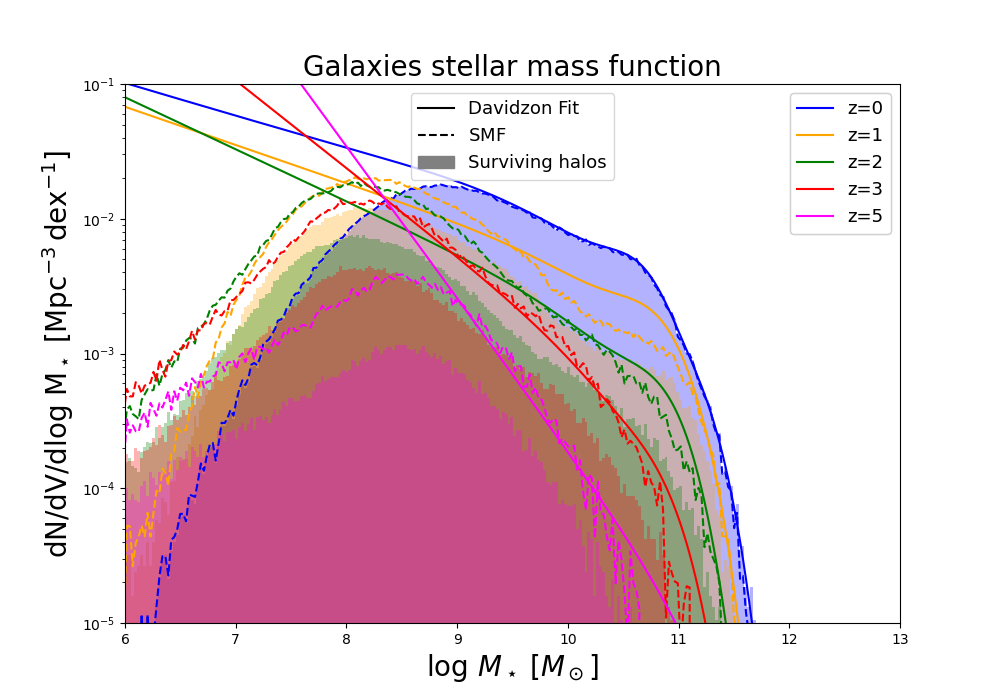}
\caption[]{Figure from \cite{Boco2023} showing the total stellar mass function evolution with redshift: blue $z=0$, orange $z=1$, green $z=2$, red $z=3$ and magenta $z=5$. Histograms represent the stellar mass function for surviving halos, dashed lines are the total mass function reconstructed from the halo catalog and solid lines show the fit by Davidzon et al. (2017) that should be considered valid for $M_\star\gtrsim$ a few $10^9\, M_\odot$. In this stellar mass range, \texttt{TOPSEM} is able to well reproduce the evolution of stellar mass function up to $z\sim 5$.}
\label{fig|SMF_model}
\end{figure*}

Despite its simplicity, \texttt{TOPSEM} is able to reproduce reasonably well the evolution of the stellar population in galaxies, as shown in Figure \ref{fig|SMF_model} which displays the comparison between the stellar mass function for surviving halos in our catalog (histograms), the predicted stellar mass function for all the galaxies (dashed lines), and the \cite{Davidzon2017} fit to observations (solid lines; this should be considered valid for $M_\star\gtrsim$ a few $10^9\, M_\odot$) at different redshifts $z$ (color code). Since at $z\approx 0$ our initial condition assigns stellar masses on the basis of the SMHM, the local stellar mass function is reproduced by construction. However, \texttt{TOPSEM} is able to fairly trace its evolution at all redshifts up to $z\sim 5$, where it starts lacking statistics at high stellar masses. The good match with the observed global stellar mass function at all redshifts further supports the assumption of monotonicity between sSFR and sHAR.

\begin{figure}[t]
\centering
\includegraphics[width=.49\textwidth]{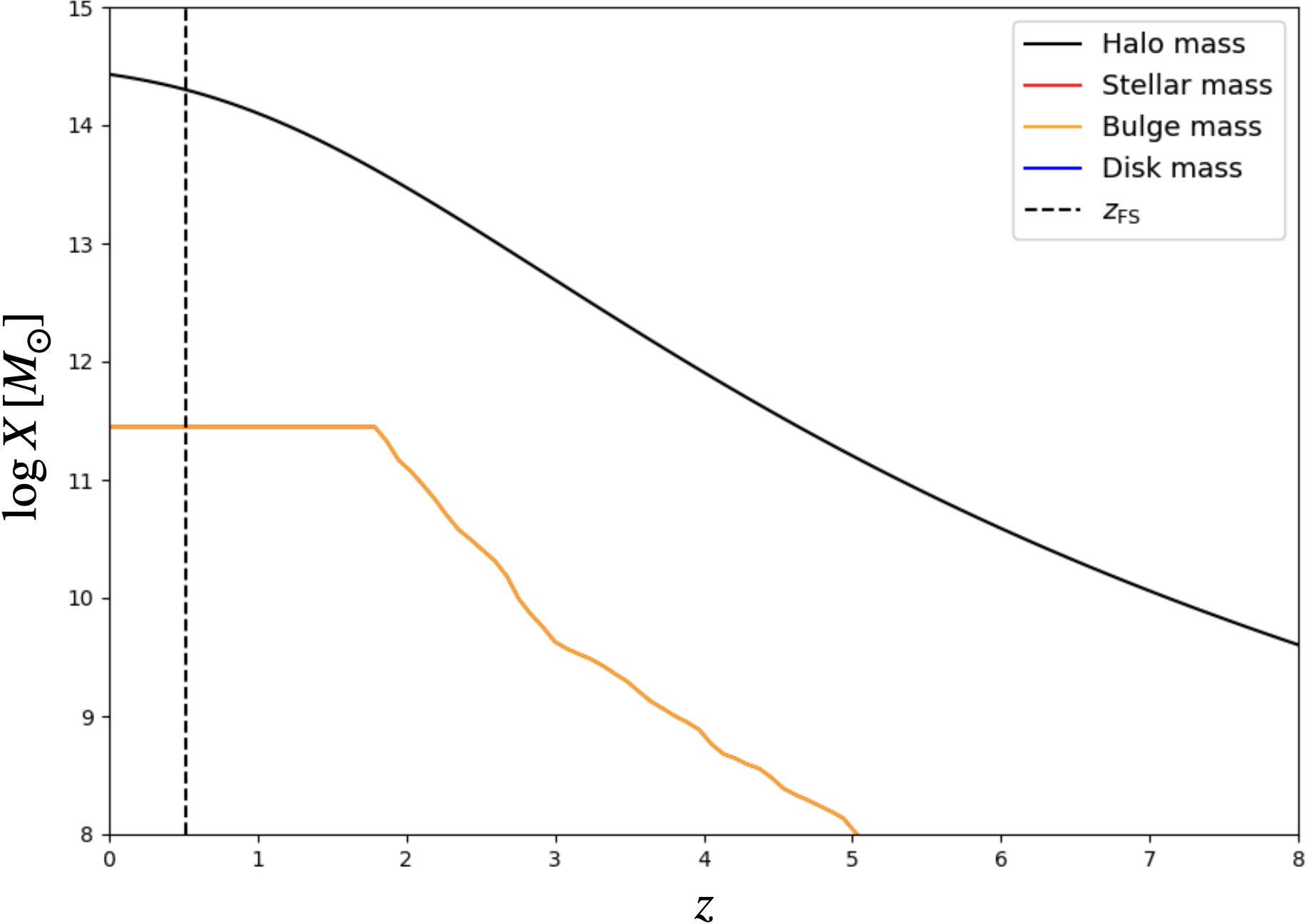}~~~~~~~~~
\includegraphics[width=.49\textwidth]{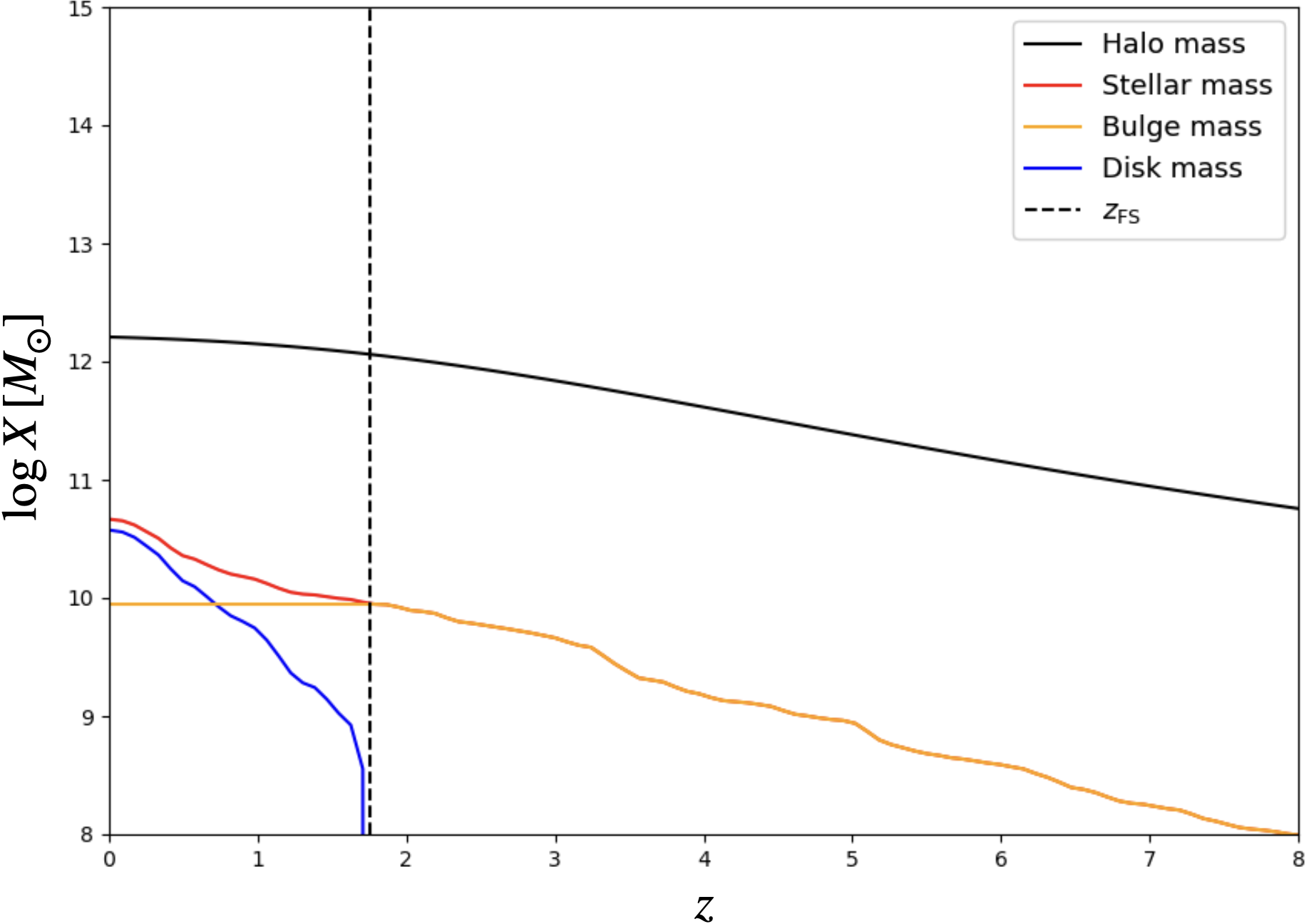}
\caption{Figure from \cite{Boco2023} showing examples of evolution for bulges and disks for two galaxies in \texttt{TOPSEM}. Black line: halo mass, red line: total stellar mass, orange line: bulge mass, blue line: disk mass. Dashed vertical line represents the transition epoch between fast and slow accretion $z_{\rm FS}$. Left panel shows a quenched galaxy with $z_Q>z_{\rm FS}$, all the stellar mass in the bulge and the galaxy is an elliptical. Right panel shows a star-forming galaxy building its bulge up to $z_{\rm FS}\simeq 1.5$ and its disk at lower redshift.}
\label{fig|evolution_bd}
\end{figure}

\begin{figure}[t]
\centering
\includegraphics[width=0.5\textwidth]{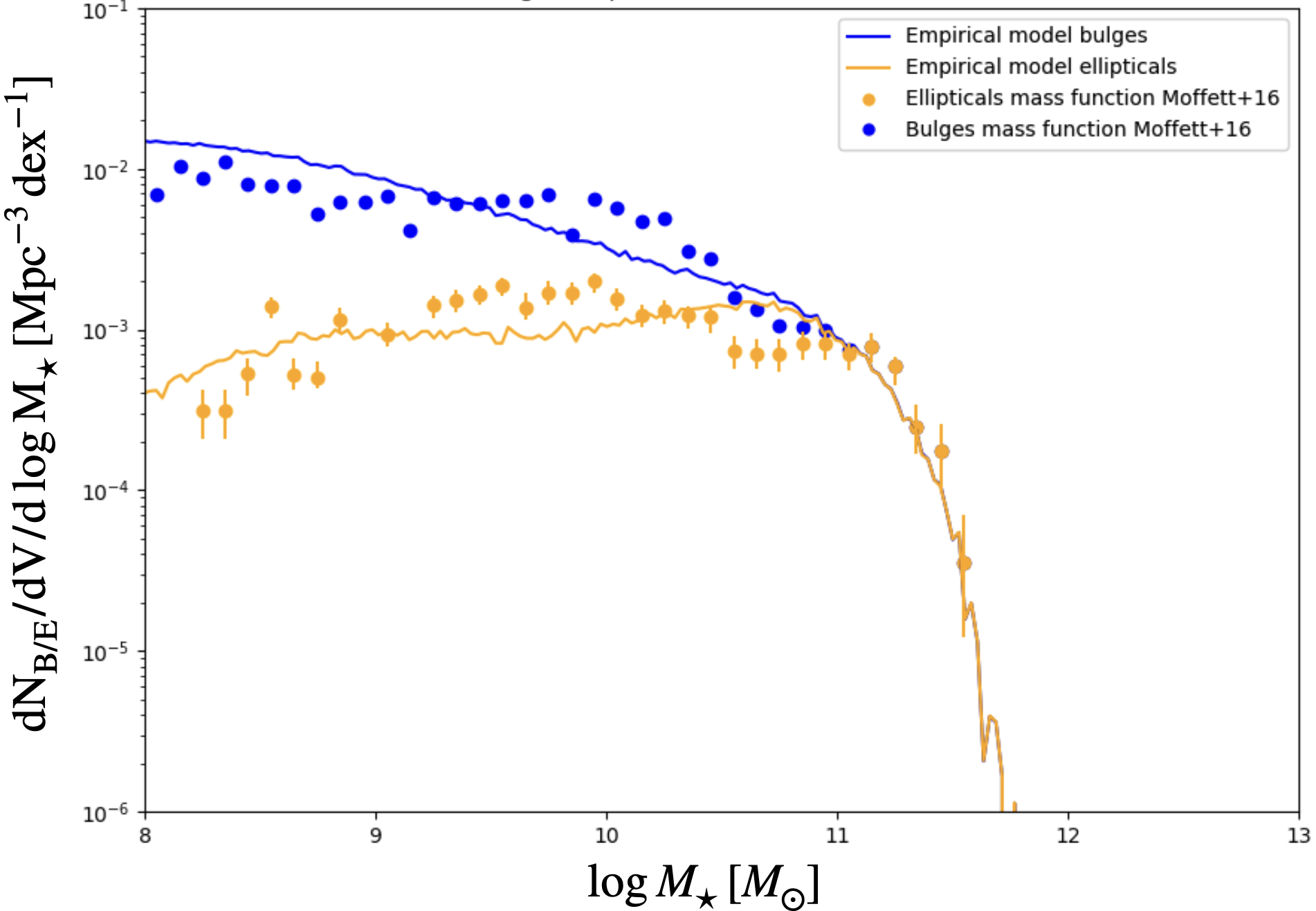}~~~~~~~~~
\includegraphics[width=.47\textwidth]{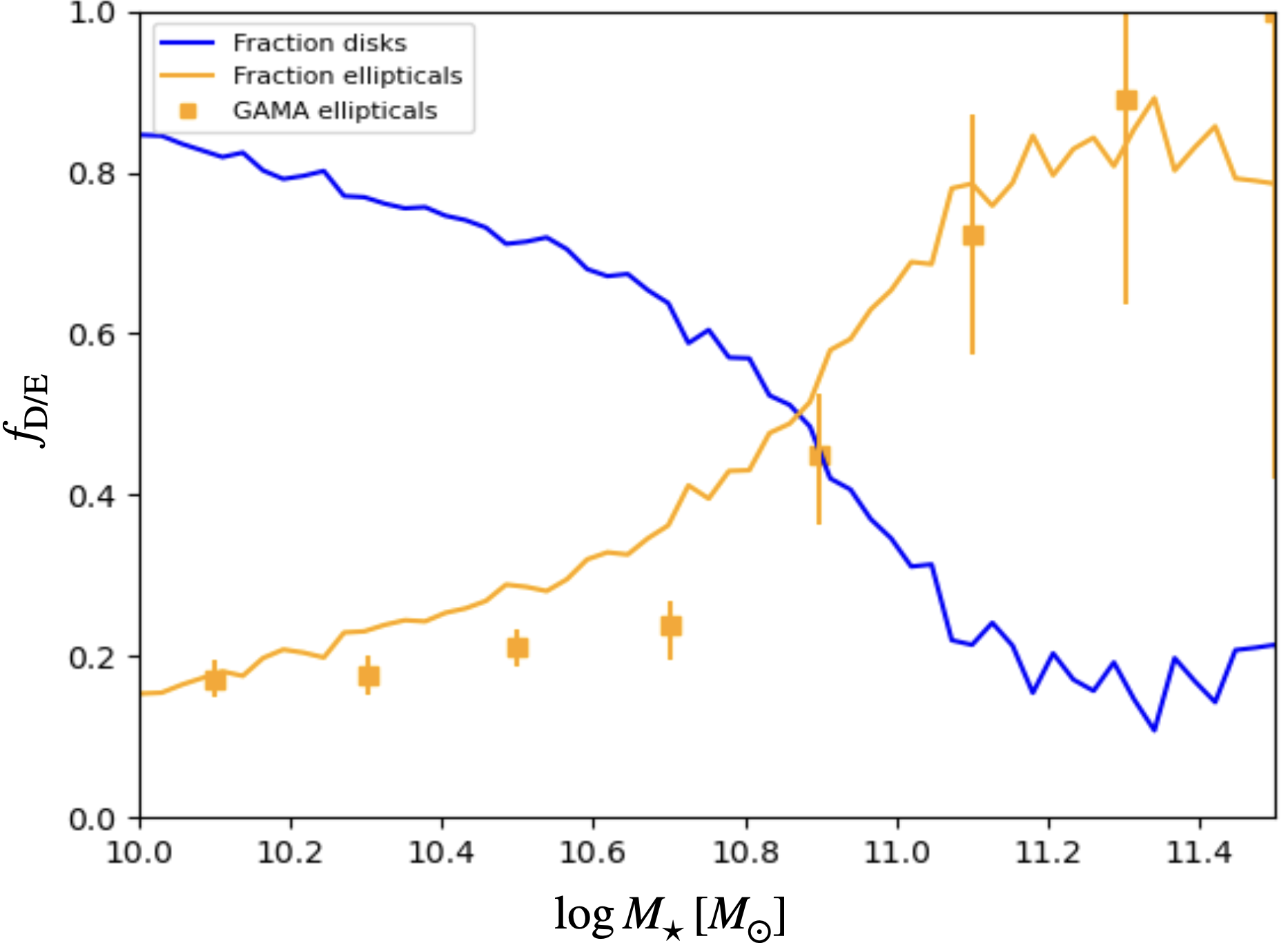}
\caption{Figure from \cite{Boco2023} showing the outcomes of \texttt{TOPSEM}. Left panel: bulge stellar mass function. Solid lines are model predictions, dots are GAMA data by \cite{Moffett2016b,Moffett2016a}. In blue the total bulge stellar mass function, in orange the stellar mass function for ellipticals. Right panel: fraction of disks and ellipticals as a function of stellar mass at $z=0$. Solid lines are the model results, square dots display GAMA data by \cite{Moffett2016b,Moffett2016a}. Orange line refers to the ellipticals fraction, while the blue line to the disk fraction.}
\label{fig|SMF_B}
\end{figure}

\subsection{Testing hypothesis on morphology}\label{sec|morphology}

In the previous Sections we have described how \texttt{TOPSEM} constructs a galaxy catalog by connecting halo and galaxy properties: this is the \emph{descriptive} component of the model. Here we show how this handy model can be exploited to test a specific hypothesis for the morphological evolution of galaxies: this is the \emph{interpretative} component of the model.  The idea is extremely simple: the galactic bulge and spheroids are formed during the fast phase of DM halo growth, while disks are originated in the subsequent slow accretion phase. 
On this basis, the amount of stellar mass in the bulge $M_{\star,b}$ and in the disk $M_{\star,d}$ can be predicted for every galaxy in the catalog and directly depends on the galaxy star formation history and on the transition redshift between fast and slow accretion $z_{\rm FS}$ of the halo. There are three possible occurrences, which depend on the relative position of the transition redshift $z_{\rm FS}$ and the quenching redshift $z_Q$:
\begin{itemize}
\item $z_{\rm FS}<0$. There are some cases in which halos, especially massive ones, are still in the fast accretion regime at $z=0$. In these cases all the stellar mass formed is in the bulge, i.e., the galaxy is an elliptical.
\item $z_{\rm FS}>0$ and $z_Q>z_{\rm FS}$. In this case the transition between fast and slow accretion occurs before $z=0$, but the galaxy has already quenched before the transition. Also in this case all the stellar mass formed is in the bulge and the galaxy is an elliptical.
\item $z_{\rm FS}>0$ and $z_Q<z_{\rm FS}$. In this case the transition between fast and slow accretion occurs before $z=0$ and quenching happens after the transition or does not happen at all. In this case all the stellar mass formed before the transition is in the bulge and the rest in the disk. The relative abundance of bulge and disk components are determined by the precise values of $z_Q$ and $z_{\rm FS}$: an early transition and no quenching would result in a galaxy with a prominent disk; a late transition, close to the quenching time, would result in a bulge-dominated/lenticular galaxy.\\
\end{itemize}

Two examples of the SFH of galaxies as a function of redshift can be seen in Figure \ref{fig|evolution_bd}. The bulge mass growth is shown in orange, the disk mass in blue and the total stellar mass in red. The left panel shows a case in which the fast-slow transition occurs after quenching; therefore all the stellar mass is in the bulge. In the right panel it is shown the case of a galaxy with no quenching, so that the separation between the bulge and disk components is evident. In this respect, \texttt{TOPSEM} can also shed light on the relation between morphology and star formation (see Section \ref{sec|intro}). As a matter of fact, elliptical galaxies are found to be more massive and quiescent, while disk galaxies are generally less massive and star-forming in the local Universe. In \texttt{TOPSEM}, quenching is not directly related to morphology, but a link between them naturally emerges. Indeed, the transition redshift $z_{\rm FS}$, on average, decreases for higher descendant halo masses. Consequently, massive halos, typically hosting massive quiescent galaxies, have less or no time to develop a substantial stellar disk and they tend to host pure ellipticals. Contrariwise, star-forming galaxies are usually hosted in lower mass halos, which have a higher transition redshift and more time to develop a stellar disk.\\

It is important now to statistically evaluate the hypothesis for disk and bulge formation comparing it with the available data for galaxy morphology. In the left panel of Figure \ref{fig|SMF_B} it is shown the $z\approx 0$ stellar mass function for bulges and ellipticals, compared with the observational GAMA data from \cite{Moffett2016b,Moffett2016a}: the agreement is impressive. The right panel, instead, shows the fraction of elliptical galaxies as a function of the stellar mass at $z\approx 0$ (orange line). This  fraction increases with $M_\star$, ranging from $\sim 0.15$ at $M_\star\sim 10^{10}\,M_\odot$ to $\sim 0.75$ at $M_\star\sim 10^{11}\,M_\odot$. Even in this case the agreement with observations is very good, at least up to $M_\star\gtrsim 10^{11}\,M_\odot$. At higher stellar masses the fraction of ellipticals remains high but the trend in the model outputs becomes unclear. This is due at least to two reasons: (i) lack of statistics, i.e., at high stellar masses the number of galaxies per mass bin becomes very low (of the order of unity) creating a large scatter on the results; (ii) dry mergers may become important and contribute in shaping the morphology of galaxies. All in all, these results indicate that the hypothesis linking the bulge/disk dicothomy to different halo accretion modes is very promising.\\

\section{What next?}\label{sec|discussion}

In this Section we highlight a few aspects that have only started being considered in semi-empirical approaches and
could represent future avenues of exploration or of application for next-generation SEMs.\\

\begin{itemize}

\item \emph{Coevolution of galaxies and supermassive black holes}

The role in galaxy evolution of (super)massive black holes (BHs) with
masses $M_\bullet\sim 10^{6-10}\, M_\odot$ constitute a crucial yet long-standing problem in modern astrophysics and cosmology. These monsters are thought to have grown mainly by gaseous accretion onto a disk surrounding the BH (e.g., \citealt{LyndenBell1969,Shakura1973}) that energizes the spectacular broadband emission of active galactic nuclei (AGNs) and leaves a BH relic ubiquitously found at the center of massive galaxies in the local universe (e.g., \citealt{Kormendy2013}). This paradigm has received a smoking gun confirmation by the \cite{EHT2019,EHT2022} via the imaging of the BH shadow caused by
gravitational light bending and photon capture at the event
horizon of M87 and Sgr A$^\star$.
Observed tight relationships between the relic BH masses and the physical properties of the hosts, most noticeably the stellar mass or velocity dispersion of the bulge component (e.g., \citealt{Magorrian1998,Ferrarese2000,Gebhardt2000,Tremaine2002, Kormendy2013,McConnell2013,Reines2015,Shankar2016,Sahu2019,Zhu2021}) suggest that the BH and the bulge stellar mass must
have co-evolved over comparable timescales (see review by \citealt{Alexander2012}),  possibly determined by the energy feedback from the BH itself on the gas/dust content of the host during accretion episodes (see \citealt{Tinsley1980,Silk1998,Fabian1999,King2005,Shankar2006,Lapi2006,Lapi2014,Lapi2018}; for a review, see \citealt{King2015rev}). 
Moreover, the discovery of an increasing number of active BHs with masses $M_\bullet \gtrsim 10^9\, M_\odot$ at high redshifts $z\gtrsim 7$ (e.g., \citealt{Mortlock2011, Wu2015,Venemans2018,Banados2018,Reed2019,Matsuoka2019,Yang2021,Wang2021}; for a recent review see \citealt{Fan2023}), when the age of the universe was shorter than $0.8$ Gyr, has rekindled the attention on mechanisms able to rapidly produce heavy BH seeds of $10^{3-5}\, M_\odot$, thus reducing the time required to attain the final masses by standard disk accretion (see \citealt{Natarajan2014,Madau2014super,Mayer2019,Inayoshi2020,Boco2020,Boco2021b,Volonteri2021,Pacucci2024}). 
\\

In terms of SEMs, early on the issue was addressed via a pioneering model put forward by \cite{Hopkins2006}, that exploits specific AGN light curves folded with data-driven prescriptions for galaxy statistics to test hypotheses on quasar lifetimes and Eddington ratio distributions (see also \citealt{Hopkins2009}). In recent years, the focus of SEMs on the galaxy-BH coevolution issue has been rekindled with a development of the \texttt{UNIVERSEMACHINE} model called \texttt{TRINITY} (see \citealt{Zhang2023,Zhang2024}). This descriptive SEM attempts to self-consistently infer the statistical connection between DM halos, galaxies, and supermassive BHs across cosmic times by constraining an admittedly large number of parameters (of order $60$) with several galaxy and BH observables. \texttt{TRINITY} provides predictions for the buildup of BHs from early epochs to the present time, for the redshift-dependent relationships between BH and galaxy observables, for the clustering properties of galaxies and AGNs, etc. However, the issue about the formation of supermassive BHs and their evolution with the host galaxies is far from being solved, both from an observational and from a modeling perspective. On the observational side, it is worth mentioning two recent JWST discoveries. The first concerns the identification of some BHs that are overmassive with respect to the host galaxies at very high redshift $z\gtrsim 10$ (e.g., \citealt{Goulding2023,Bogdan2024,Maiolino2024}). The second involves an overmassive, X-ray underluminous BH population at intermediate $z\sim 4-8$ (e.g., \citealt{Ubler2023,Furtak2023,Harikane2023,Kokorev2023,Stone2024}), mostly located in the so called `little red dots' sources (see \citealt{Kocevski2023,Matthee2024}), that could represent (lightly) dust-reddened AGNs (see \citealt{Greene2024,Yue2024,Maiolino2024lrd}); note that their number densities tends to overproduce the X-ray luminosity function at the same redshifts by up to an order of magnitude. On the modeling side, it would be extremely valuable to explore these new exciting, but also puzzling, data through the lens of interpretative and hybrid SEMs. These can verify for possible internal self-consistency among distinct datasets, whilst testing some underlying physical scenarios via flexible and transparent approaches characterized by a small numbers of parameters. Hopefully, such approaches could lead to understand better how to form overmassive BHs in the early Universe and to clarify the role of BHs in galaxy evolution.\\

\begin{figure}
    \centering
    \includegraphics[scale=0.9]{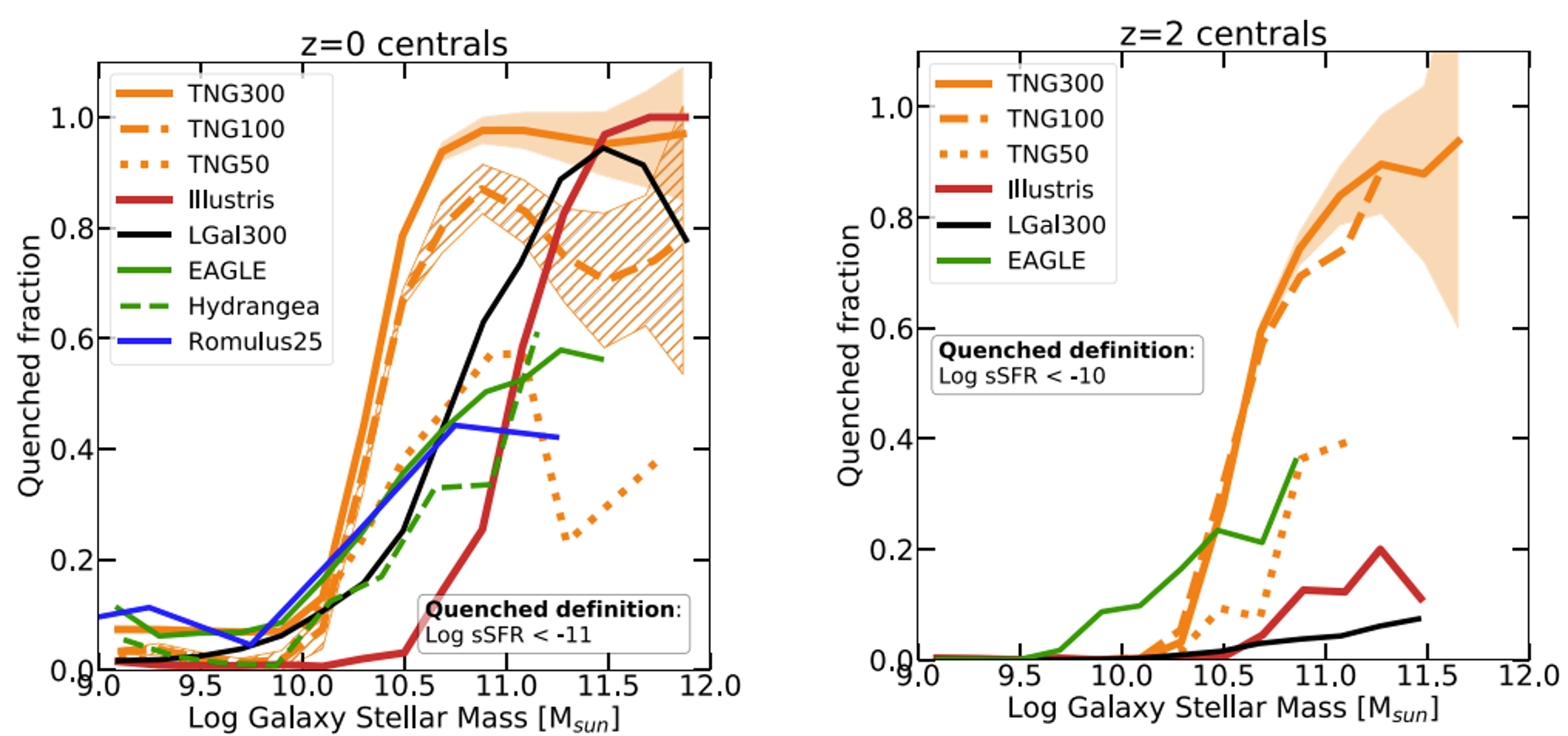}
    \caption{Figure from \cite{Donnari2021} showing the outcomes from different theoretical models of galaxy evolution for the fraction of quiescent galaxies as a function of stellar mass at $z=0$ (left panel) and $z=2$ (right panel).}
    \label{fig|illustris}
\end{figure}

\item \emph{Galaxy quenching mechanisms}

One of the most important and mysterious issues in galaxy evolution concerns how galaxies quench their star formation. Proposed mechanisms include: morphological quenching, possibly related to the presence of the stellar bulge (e.g., \citealt{Cook2009,Martig2009,Gensior2020}); negative energy/momentum feedback via stellar winds, supernova explosions and accreting supermassive BHs that can heat and eject gas from starforming regions (e.g., \citealt{Silk1998,Fabian1999,Granato2004,King2005,Lapi2006,Grand2017,Lapi2018,Valentini2020,Goubert2024}); mass quenching via a threshold in halo mass above which galaxies can retain a hot gas atmosphere that hinder further cold gas inflows (e.g., \citealt{Birnboim2003,Keres2005,Dekel2006,Dekel2009,Dekel2023}); environmental quenching via tidal and ram-pressure stripping of the gas during interactions among galaxies or between a galaxy and the diffuse environment (e.g., \citealt{Larson1980,Moore1999,Bekki2002, Roediger2014,Steinhauser2016,Man2018,Mao2022}). 
Reasonably, all these effects could combine to various degree in galaxies with different states and environments. As of now, there is no definitive observational evidence pinpointing the leading mechanism responsible for quenching galaxies of different morphological types. Even from a theoretical perspective, ab initio approaches (like SAMs and SIMs) struggle somewhat in highlighting the primary physical mechanism responsible for quenching galaxies, and their outputs are often controversial, degenerate and strongly dependent on detailed physical recipes (e.g., \citealt{Scannapieco2012,Donnari2021,Habouzit2022,Crain2023}). As an example, Figure \ref{fig|illustris} shows the predictions of different theoretical models of galaxy evolution for the fraction of quiescent galaxies at different stellar masses and redshifts $z\approx 0$ (left panel) and $z\approx 2$ (right panel). It appears evident that accurately reproducing the observed number density and properties of quiescent galaxies at different cosmic epochs remains an unsolved problem.\\

In this complex landscape, SEMs of galaxy evolution could play a pivotal role in 
setting more stringent constraints on the origin of quenching. The most effective route could be to design a hybrid SEM where specific quenching hypotheses are applied on top of a very basic data-driven framework, gauged on the observed statistical properties of galaxies in different states (for a first attempt see, e.g., \citealt{Fu2025}). Then each of the putative triggers for quenching can be systematically tested, in isolation or in combination with others, against the observed abundances and properties of quenched galaxies at different cosmic epochs. In this way the modeling would avoid becoming dependent on too many parameters and being exposed to the danger of serious degeneracies or divergences. The impact of any physical hypothesis for quenching could thus be disentangled clearly in the outcomes. Thanks to their flexibility and transparency, SEMs could also point/reveal new routes to test the input assumptions on quenching. 
\\

\item \emph{Synergy with data science techniques}

The advent of data science has revolutionized the field of galaxy evolution, making it possible to handle, store, mine and analyze the vast amounts of outputs associated to present and upcoming observational facilities and numerical studies. The related datasets are intrinsically rich in volume, variety, and often include information that cannot be easily described via traditional Bayesian likelihoods approaches. To face these challenges, astrophysicists have started relying on data-driven, machine learning, and AI-enhanced solutions; many of these could interface very well with the SEM spirit. For example, ensemble sampling algorithms like affine Markov Chain Monte Carlo (e.g., \citealt{Goodman2010,Foreman2013}), Hamiltonian Monte Carlo (e.g., \citealt{Neal2011,Betancourt2013}), Dynamic Nested sampling (e.g., \citealt{Skilling2004,Higson2019}) allow to efficiently explore the parameter space of a SEM, to optimize the agreement between the parametric model with the target datasets, and to provide a robust assessment of parameter uncertainties and degeneracies. On a more advanced perspective, AI-enhanced algorithms like deep learning (e.g., \citealt{Huertas2023}), generative adversarial networks (e.g., \citealt{Goodfellow2014}), variational autoencoders (e.g., \citealt{Kingma2013}), normalizing flows (e.g., \citealt{Dinh2014}), etc. based on SEM outputs could be devised to connect galaxies and DM halo properties, and then exploited to quickly populate large-volume SIMs (see \citealt{Kamdar2016a,Kamdar2016b,Agarwal2018,Hassan2022,Appleby2023}), or even to emulate sub-grid physics in SIMs via generation of super-resolution tracer particles (e.g., \citealt{Li2021,Zhang2024DS}).\\

\end{itemize}

\section{Summary}\label{sec|summary}

Modeling galaxy formation and evolution is admittedly a very difficult task since galaxies constitute extremely complex systems, involving many processes over a huge range of time, space and energy scales. Although the precision of the astrophysical measurements are in no way comparable to other fields of ‘experimental’ physics, the amount of data is extraordinarily rich and the ensuing constraints on theoretical models are overwhelming. 
Solving open issues of galaxy formation and evolution therefore requires complementarity among different modeling approaches:  (semi-)analytic, semi-empirical, and numerical models must be jointly exploited to refine sub-resolution physics, to choose the best parameterizations of the different processes under study, and to understand the inconsistencies and biases among observables. \\

In this review, we have provided a census of semi-empirical models (SEMs) of galaxy formation and evolution, stressing their value in being data-driven, easily expandable, and computationally low-cost approaches. We have highlighted different flavors of SEM, i.e. interpretative, descriptive and hybrid, discussing the peculiarities, virtues and shortcomings in each of these variants. We have also dissected 
a recent hybrid SEM, to describe in handy terms the techniques employed by these data-driven approaches to test simple hypotheses by marginalizing over unknown processes and avoiding  heavy parameterizations. Finally, we have provided an outlook on possible future developments and applications of SEMs, related to the role of supermassive black holes in galaxy evolution, to galaxy quenching mechanisms, and to the synergy with advanced data science techniques.

\begin{ack}[Acknowledgments]

We warmly thank L. Danese, H. Fu and the GOThA team for useful discussions and helpful suggestions. This work is partially funded from the projects: ``Data Science methods for MultiMessenger Astrophysics \& Multi-Survey Cosmology'' from the Italian Ministry of University and Research, Programmazione triennale 2021/2023 (DM n.2503 dd. 9 December 2019), Programma Congiunto Scuole; EU H2020-MSCA-ITN-2019 n. 860744 \textit{BiD4BESt: Big Data applications for Black hole Evolution Studies} (coordinator F. Shankar); Italian Research Center on High Performance Computing Big Data and Quantum Computing (ICSC), project funded by European Union - NextGenerationEU - and National Recovery and Resilience Plan (NRRP) - Mission 4 Component 2 within the activities of Spoke 3 (Astrophysics and Cosmos Observations); European Union - NextGeneration EU, in the framework of the PRIN MUR 2022 project n. 20224JR28W "Charting unexplored avenues in Dark Matter"; INAF Large Grant 2022 funding scheme with the project "MeerKAT and LOFAR Team up: a Unique Radio Window on Galaxy/AGN co-Evolution; INAF GO-GTO Normal 2023 funding scheme with the project "Serendipitous H-ATLAS-fields Observations of Radio Extragalactic Sources (SHORES)".
\end{ack}

\seealso{article title article title}

\bibliographystyle{Harvard}
\bibliography{reference}

\end{document}